\documentclass[aps,pre,superscriptaddress,a4paper,showpacs]{revtex4}
\usepackage{graphicx,amsmath,amssymb,psfrag,curves,epic,subeqnarray}
\usepackage{natbib}
\usepackage{easymat}
\usepackage{easybmat}
\usepackage{enumerate}
\newcommand{\bx}{\mathbf x} 
\newcommand{\bU}{\mathbf U} 

\def\be{\begin{equation}}
\def\ee{\end{equation}}
\def\tdelh{\frac{\Delta t}{2}}
\def\tdel{\Delta t}
\def\lap{\nabla^2}
\def\divv{\grad\cdot}
\def\bnab{{\mathbf \nabla}}
\def\zdir{{\mathbf e_z}}
\def\grad{\bnab}

\def\lef{\left(}
\def\rig{\right)}

\def\etal{{\em et al.}}

\def\Hmax{\bar{H}}

\begin{document}

\title{Extreme multiplicity in cylindrical Rayleigh-B\'enard convection:\\
I.~Time-dependence and oscillations}

\author{Katarzyna Boro\'nska}
\email[]{k.boronska@leeds.ac.uk}
\homepage[]{www.comp.leeds.ac.uk/kb}
\author{Laurette S.\ Tuckerman}
\email[]{laurette@pmmh.espci.fr}
\homepage[]{www.pmmh.espci.fr/~laurette}
\affiliation{Laboratoire d'Informatique pour la M\'ecanique et les Sciences 
de l'Ing\'enieur (LIMSI­-CNRS), B.P. 133, 91403 Orsay, France}

\begin{abstract}
Rayleigh-B\'enard convection in a cylindrical container can take on many
different spatial forms. Motivated by the results of Hof, Lucas and Mullin [{\em
    Phys.~Fluids\/} {\bf 11}, 2815 (1999)], who observed coexistence of
several stable states at a single set of parameter values, we have carried out
simulations at the same Prandtl number, that of water, and 
radius-to-height aspect ratio of two.  We have used two kinds of thermal
boundary conditions: perfectly insulating sidewalls and perfectly conducting
sidewalls.  In both cases we obtain a wide variety of coexisting steady and
time-dependent flows.
\end{abstract}

\date{\today}
\maketitle

\section{Introduction}

\begin{figure}[!htbp]
\begin{center}
    \includegraphics[scale=8]{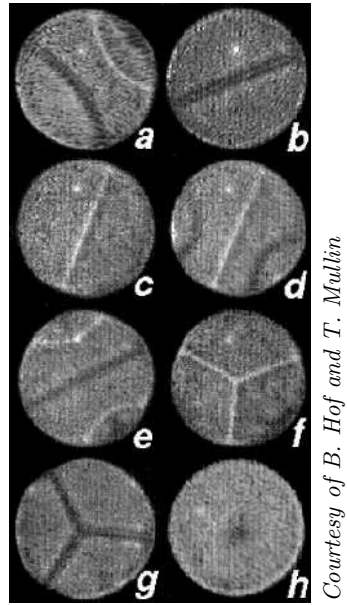}\hspace{5pt}\rotatebox{90}{{\it Courtesy of B. Hof and T. Mullin}}
\end{center}
     \caption{Patterns observed in the experiment of Hof \etal at
       $Ra=14\,200$.  (\textit{a}) three rolls, (\textit{b}) two rolls,
       (\textit{c}) inverted two rolls, (\textit{d}) four rolls, (\textit{e})
       inverted four rolls, (\textit{f}) mercedes, (\textit{g}) inverted
       mercedes, (\textit{h}) axisymmetric pattern.  Dark areas correspond to
       hot (rising) and bright to cold (descending) fluid.}
    \label{fig:hof}
\end{figure}

The flow patterns realized by Rayleigh-B\'enard convection depend not only on
the fluid parameters and container geometry, but also on the flow history.
This was strikingly illustrated by the experimental study carried
out by Hof \etal{}~\cite{HofLucMul,HofThesis}. They investigated a cylinder of small
aspect ratio  $\Gamma\equiv{\rm radius/height}=2.0$ filled with water
(Prandtl number 6.7).  
Varying the Rayleigh number through different sequences of values, 
they obtained several
different stable patterns for the same final Rayleigh number. For $14\,200$
they observed two, three and four parallel rolls, a ``mercedes'' pattern with
three spokes of ascending or descending fluid and even an axisymmetric state (figure~\ref{fig:hof}).
They also reported a transition from an axisymmetric steady state towards
azimuthal waves.

We begin by reviewing the literature on Rayleigh-B\'enard convection in
cylinders with small-to-moderate aspect ratio $1\lesssim\Gamma\lesssim 10$.
Our interest in such geometries stems from the fact that
they are the battleground of two competing tendencies
over an appreciable Rayleigh-number range.
At threshold, the patterns necessarily resemble eigenmodes of the Laplacian, 
i.e.~they are trigonometric in the azimuthal angle.
But at higher $Ra$, the patterns form rolls, bending and pinching so 
as to fit into the cylinder. 

The instability of the conductive state in a cylindrical geometry
was well established in the 1970s--1980s 
\cite{ChaSan70,ChaSan71,StoMul,BueCat}.  Critical Rayleigh numbers
$Ra_c$ are about $2000$ for $\Gamma\sim 1$, increasing steeply for lower
$\Gamma$ and decreasing asymptotically towards $Ra_{c}=1708$ for
$\Gamma\rightarrow\infty$.
The seminal paper of Buell and Catton~\cite{BueCat} surveyed the influence of 
sidewall conductivity on the onset of convection by performing linear
analysis for the aspect ratio range $0 < \Gamma \leq 4$.
They determined the critical Rayleigh number and azimuthal wavenumber as a 
function of both aspect ratio and sidewall conductivity, thus completing the 
results of the previous investigations, which considered either perfectly 
insulating or perfectly conducting walls. The flow succeeding the conductive 
state is three-dimensional over large ranges of aspect ratios, contrary to the 
expectations of Koschmieder~\cite{Kos}. Similar calculations were 
carried out by Marques \etal~\cite{Marques93}.

The stability of the first convective state,
for situations in which the primary flow is axisymmetric,
was investigated in the 1980s--1990s~\cite{MulNeuWeb,HarSan,WanKuhRat,TouHadHen} 
for a variety of aspect ratios and Prandtl numbers.
Bifurcation-theoretic scenarios for axisymmetric flows have been
investigated by Barkley and Tuckerman~\cite{BarTuc} and 
Siggers~\cite{Sig}.
Experiments on the competition between various convective patterns
in a constrained cylindrical geometry ($\Gamma \approx 7.5$) were 
carried out during this period~\cite{PocCroLeG,Cro,SteAhlCan} 
and the dynamics interpreted in terms of the instabilities 
of two-dimensional straight rolls \cite{Busse79}.

Several time-dependent simulations since 2000 have focused on the variety of
coexisting non-axisymmetric patterns for $\Gamma \approx 4$.  R{\"u}diger and
Feudel~\cite{RudFeu} used a spectral simulation to investigate the aspect
ratio $\Gamma=4$, finding the stability ranges of several patterns -- parallel
rolls, target and spirals -- with overlapping stability ranges in Rayleigh
number.  Paul \etal{}~\cite{PauChiCroFisGre} studied the dislocation dynamics
of curved rolls near onset.

Several studies have focused on configurations similar or identical to that of
Hof \etal{}~\cite{HofLucMul}.  Leong~\cite{Leong} used a finite difference
method to simulate convective flows for Prandtl number $Pr=7$ in cylinders of
aspect ratios $\Gamma=2$ and $4$ and adiabatic lateral walls and observed
several steady patterns -- parallel rolls, three-spoke flow and axisymmetric
state -- all of which were stable in the range $6250\leq Ra\leq 37\,500$.  
Ma \etal{}~\cite{Ma} used a finite volume method to simulate the parameters 
$\Gamma=2$ and $Pr=6.7$ of Hof \etal{}~\cite{HofLucMul}.
As in our prior study \cite{Boronska_PhD},
they used time-dependent simulation at $Ra=14\,200$
to reproduce the five states of Hof \etal{} 
They also calculated the first nine primary bifurcations 
to modes of varying azimuthal wavenumbers, 
and several secondary bifurcations, all taking place for $Ra\leq 2500$.
However, they reported no results for $Ra$ between 2500 and 14\,200,
nor for any values above 14\,200.

The work we present here is based on that of Boro\'nska~\cite{Boronska_PhD} and
covers simulations of $\Gamma=2$ and $Pr=6.7$ over
the entire Rayleigh-number range up to $30\, 000$.
We carry out a complete survey of the convective patterns
we locate; we then carry out the same study for
the case of conductive sidewalls. 
A companion paper \cite{Boronska_PRE2} takes the present 
article as a starting point to compute a bifurcation diagram 
for the insulating sidewall case
over the range $Ra \leq 30\,000$, which connects the patterns 
at $Ra = 14\,200$ with the primary bifurcations from the 
conductive branch.

\section{Governing equations and numerical methods}

The methods used in our study are described in this section.
We first present the nonlinear equations governing the 
system. We then
describe the most important aspects of the numerical methods used for
integrating the differential equations. 
A detailed description of all numerical techniques used can be found in 
Tuckerman~\cite{Tuc}.

\subsection{Equations and boundary conditions}
\label{sec:equations}

\begin{figure}[!htbp]
\begin{center}
    \includegraphics[width=0.5\textwidth]{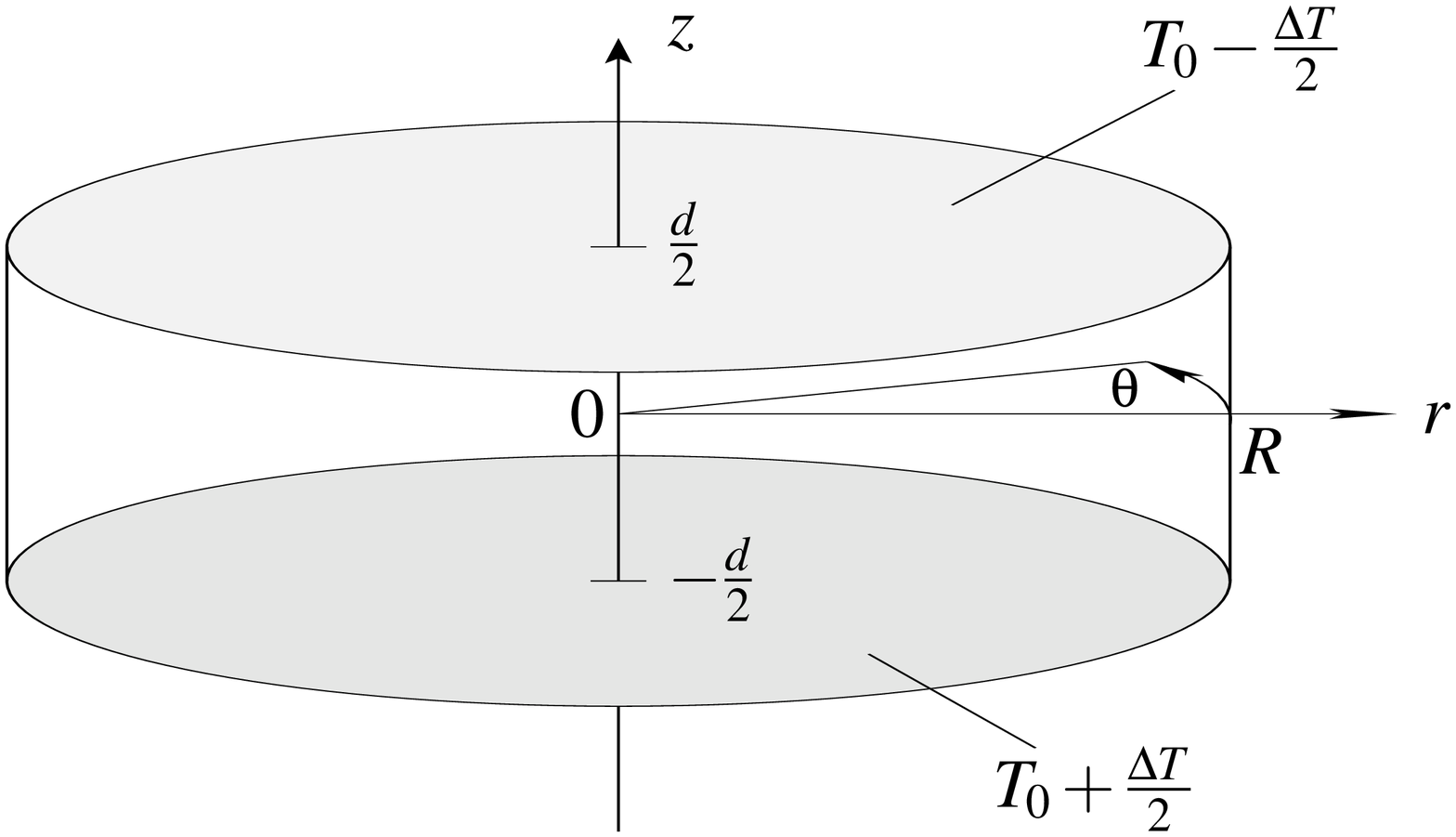}
\end{center}
     \caption{Geometry and coordinate system.}
    \label{fig:cylinder}
\end{figure}

We consider a fluid confined in a cylinder of depth $d$ and radius $R$,
whose aspect ratio is defined as $\Gamma\equiv R/d$ (figure~\ref{fig:cylinder}).  
The fluid has kinematic viscosity $\nu$, density $\rho$, thermal diffusivity $\kappa$ and
thermal expansion coefficient (at constant pressure) $\gamma$. The top and
bottom temperatures of the cylinder are kept constant, at $T_0-\Delta T /2$
and $T_0+\Delta T /2$, respectively. 
The Rayleigh number $Ra$ and the Prandtl number $Pr$ are defined by
\begin{equation}
Ra \equiv \frac{\Delta T g \gamma d^3}{\kappa\nu}, \qquad\qquad
Pr \equiv \frac{\nu}{\kappa}.
\end{equation}
Using the units $d^2/\kappa$, $d$, $\kappa/d$ and $\nu \kappa /\gamma g d^3$
for time, distance, velocity and temperature, and defining $H$ 
to be the dimensionless temperature deviation from the linear vertical profile,
we obtain the Navier--Stokes and
Boussinesq dimensionless equations governing the system:
\begin{subequations}
\begin{eqnarray}
\partial _t H
+\left(\mathbf U\cdot\nabla\right)H &=& Ra\;U_z+\lap H \label{eq:N-Sa}\\
\label{eq:N-S}
Pr^{-1}\left({\partial _t}\mathbf{U} +
\left(\mathbf{U}\cdot{\mathbf\nabla}\right)\mathbf{U}
\right)
 &=& - {\mathbf\nabla}P + \lap\mathbf{U}
+H{\mathbf e_z} \label{eq:N-Sb}\\
{\mathbf\nabla}\cdot\mathbf{U}&=&0
\label{eq:N-Sc}
.
\end{eqnarray}
\end{subequations}
We used realistic boundary conditions for the velocity, with 
no-penetration and no-slip 
on the horizontal plates and sidewalls:
\begin{subequations}
\begin{equation}
{\bf U} = 0 \qquad \quad {\rm for} \quad  z=\pm 1/2 \quad {\rm or} 
\quad r=\Gamma. 
\label{eq:bcu}\end{equation}
We assume that perfectly conducting horizontal plates maintain the temperature
constant (homogeneous Dirichlet condition for $H$)
\be
	H= 0 \quad{\rm for}\quad z=\pm 1/2. \label{eq:bchz}
\ee
The sidewalls are either perfectly insulating
(Neumann boundary conditions)
\be
	\partial_r H= 0 \quad {\rm for}\quad r=\Gamma. \label{eq:bcneu}
\ee
or perfectly conducting,
so that a linear vertical temperature profile is maintained within them
(Dirichlet boundary conditions)
\be
	H= 0 \quad{\rm for}\quad r=\Gamma. \label{eq:bcdir}
\ee
\label{eq:bcs}
\end{subequations} 
The code has also a possibility we have not utilized here, of interpolating
between these two conditions in order to represent sidewalls with finite
values of conductivity, as was done in \cite{BarTuc}:
\be
\mu_{Dir} H + \mu_{Neu} \partial _r H = 0\quad{\rm for}\quad r=\Gamma .
\ee
\subsection{Spatial and temporal discretization}

We represent the fields by a tensor product of 
trigonometric functions in $\theta$ and Chebyshev polynomials 
on the intervals $-1/2\leq z \leq 1/2$ and $0\leq r \leq \Gamma$.
A representation which is regular at the origin can be created by imposing
parity restrictions on the Chebyshev and Fourier functions
\cite{Tuc}. The temperature and vertical velocity are approximated
by the formula
 \begin{equation}
f(r,\theta,z)=\sum^{N_\theta/2}_{m=0}
\sum^{2N_r-1}_{{j\geq m}\atop{j+m\ \mathrm{even}}}
\sum^{N_z-1}_{k=0}\hat{f}_{j,m,k} \: T_j(r/\Gamma)\: T_k(2z) \: e^{im\theta}
+{\rm c.c.}
\label{eq:ztRep2}
\end{equation}
For the radial and azimuthal velocity components, the same representation is
used, except with $j+m$ odd.  In the code, less stringent parity rules are
used, ensuring only that the first three derivatives of $f$ are continuous at
the cylinder axis. The pseudospectral method \cite{GotOrs} calls for carrying
out differentiation on the spectral representation on the right-hand-side of
\eqref{eq:ztRep2}. Multiplication is carried out by first transforming to a
representation on a Gauss-Lobatto grid, which includes the boundary points.
The Fourier transforms, including cosine transforms for the Chebyshev 
polynomials, use the Temperton algorithms \cite{Temperton} 
for dimensions which are not restricted to powers of two.
Radial and axial boundary conditions are imposed via the tau method,
i.e.~replacing the equations corresponding to the highest order Chebyshev
polynomials by the boundary conditions.


The temporal discretization is semi-implicit. 
The viscous, diffusive, and buoyancy terms are integrated via the 
backwards Euler scheme while the advective terms are integrated via the 
Adams--Bashforth scheme.
Applying these formulas to \eqref{eq:N-Sa}-\eqref{eq:N-Sb}, we obtain:
%
%
\begin{subequations}
\begin{eqnarray}
\lef 1-\tdel \lap \rig H^{n+1}
=
H^n &+&\tdelh\left[
-  3 \lef \bU^n\cdot \grad \rig H^n + 
\lef \bU^{n-1} \cdot \grad \rig H^{n-1}
+Ra \lef 3{U_z}^n -{U_z}^{n-1} \rig \right] \label{eq:timeschema:b}\\
\left(1-\tdel \,Pr\,\lap \right) \bU^{n+1} + \tdel Pr \grad \, P^{n+1}
&=&	\bU^n+\tdelh
\left[-3 \left( \bU^n\cdot \bnab \right)\bU^n 
+ \left( \bU^{n-1}\cdot \bnab \right)\bU^{n-1}\right]
+\tdel Pr \,H^{n+1} \zdir.
\label{eq:timeschema:a}%
\end{eqnarray}
\label{eq:timeschema}
\end{subequations}

In the Helmholtz operators on the left-hand-sides of \eqref{eq:timeschema},
each Fourier mode $m$ in the representation \eqref{eq:ztRep2}
is decoupled from the other Fourier modes,
In addition to this economy, for each $m$, 
the matrix representing a differential operator can be 
diagonalized in the $z$ direction \cite{CanHusQuaZan} and 
reduced to a banded matrix in the $r$ direction \cite{Tuck_banded}.
The storage and the time for inversion of the resulting linear systems
is then on the order of $N_\theta N_r N_z^2$.
 
\subsection{Resolution}

In order to choose our spatial resolution, we carried out several 
runs at varying resolution at Rayleigh number 
$14\,000$ and timestep $\tdel = 2\times 10^{-4}$,
comparing 
the evolution of several quantities, such as temperature and velocity 
at fixed gridpoints, several spectral coefficients and the total energy
\be
E\equiv\frac{1}{Ra} \left(\frac{1}{Pr}\int_{\rm vol}{\bf U}\cdot{\bf U} + 
\frac{1}{Ra}\int_{\rm vol}H^2 \right).
\label{eq:entot}
\ee
The resolution $N_r \times N_\theta \times N_z = 40 \times 120 \times 20$
was chosen as a compromise between accuracy and efficiency.
This resolution insured that the spectral coefficients decayed 
exponentially with wavenumber or Chebyshev index.
We also reproduced a few high Rayleigh number states 
($20\,000 \leq Ra \leq 30\,000$) at the 
higher resolution $N_r \times N_\theta \times N_z = 60 \times 160 \times 30$.
In case of any uncertainty concerning the resulting pattern, 
the mesh was refined; each such test confirmed the results 
obtained with the previous resolution.
  
The timestep $\tdel$ chosen depends on Rayleigh number and on the evolution of
the system. For higher Rayleigh numbers and abrupt transitions, a smaller
timestep is required. For slowly evolving fields, the number of iterations
necessary for convergence becomes too great, unless we use larger $\tdel$.  We
chose the initial timesteps as a function of Rayleigh number and if the
evolution slowed down, we continued the simulation with an increased $\tdel$.
The timesteps we used varied from $\tdel=8\times10^{-4}$ for $Ra=2000$ to
$\tdel=2\times10^{-4}$ for $Ra=30\,000$.  The timestep was reduced when
necessary, especially when any oscillations appeared.

\subsection{Incompressibility: velocity--pressure decoupling}
In order to decouple velocity and pressure,
the divergence operator is applied to \eqref{eq:timeschema:a} to 
derive a Poisson equation for $P$
\be
\lap P = \divv \left(\frac{1}{2\, Pr}
\left[
-3 \left( \bU^n\cdot \bnab \right)\bU^n + \left( \bU^{n-1}\cdot \bnab \right)\bU^{n-1}
\right]+ H^{n+1}\zdir \right),
\label{eq:Poisson}
\ee
where $\divv U^n=\divv U^{n+1}=0$ and $\divv\lap=\lap\divv$ have been used.

The thorny issue of boundary conditions for this Poisson equation 
can be addressed in several ways \cite{Tuc,Rempfer}.
Most approaches derive boundary conditions for $P$ from the original
momentum equation \eqref{eq:N-Sb}.
This will yield a divergence which is proportional to 
a power of $\tdel$. This is usually acceptable for time integration,
where $\tdel$ is small.
However, to carry out Newton's method for steady state solving, 
as we will do in our companion paper~\cite{Boronska_PRE2}, we adapt this time-stepping
code, using $\tdel\gg 1$. We therefore require a method which
imposes incompressibility regardless of $\tdel$.
Essentially, the correct boundary condition for \eqref{eq:Poisson} is
\be
\divv \bU = 0 \qquad\qquad{\rm for} \quad z=\pm 1/2 \quad {\rm or } 
\quad r=\Gamma ,
\ee
which again couples velocity and pressure.

We use a Green's function or influence matrix method \cite{Tuc} as follows.\\
In a preprocessing step, we calculate a set of {\it homogeneous} solutions 
$\bU^{{\rm hom},j}$:
\begin{enumerate}[{\it(h,i)}]
\item We solve the homogeneous version 
of the pressure Poisson equation \eqref{eq:Poisson}
with all possible Dirichlet boundary conditions for $P$, for example,
by choosing each point $\bx_j$ on the boundary in turn and setting the 
boundary conditions to be 1 at $\bx_j$ and 0 elsewhere. (An 
equivalent approach would loop over all of the highest Chebyshev components.)
This yields a set of homogeneous pressure solutions 
$P^{{\rm hom},j}$.
\item We then solve the homogeneous version of 
\eqref{eq:timeschema:a}, i.e.
setting the right-hand-side to zero, setting 
$P=P^{{\rm hom},j}$
and imposing the homogeneous boundary conditions \eqref{eq:bcs}. 
This yields a set of homogeneous velocity solutions 
$\bU^{{\rm hom},j}$.
\item We complete the preprocessing step by calculating 
$\mathcal{C}_{ij}\equiv \divv \bU^{{\rm hom},j}(\bx_i)$ 
on all points $\bx_i$ of the boundary. $\mathcal{C}$ is the influence (or
capacitance) matrix.
\end{enumerate} 
At each timestep, we calculate a {\it particular} solution $\bU^{{\rm part}}$
and then the final solution $\bU$ as follows:
\begin{enumerate}[{\it(p,i)}]
\item We solve the pressure Poisson equation \eqref{eq:Poisson}
with the non-zero right-hand-side in \eqref{eq:Poisson},
but with homogeneous Dirichlet boundary conditions for $P$. This yields the 
particular pressure 
$P^{{\rm part}}$.
\item We solve \eqref{eq:timeschema:a} for the velocity, using 
$P=P^{{\rm part}}$
and imposing the homogeneous boundary conditions \eqref{eq:bcu}.
This yields the particular velocity 
$\bU^{{\rm part}}$.
\item We calculate 
$\divv \bU^{{\rm part}}(\bx_i)$, 
the divergence of 
the particular velocity on each point on the boundary.
\item By solving the linear system 
\be
0 = \divv \bU^{{\rm part}} (\bx_i)+ \sum_j c_j \divv\bU^{{\rm hom},j}(\bx_i)
\label{eq:infmat}\ee
involving the influence matrix, we determine the coefficients $c_j$.
\item The final solution is assembled as
\be
\bU^{n+1}=\bU^{{\rm part}}+ \sum_j c_j \bU^{{\rm hom},j}.
\ee
\end{enumerate}
Alternatively, if the homogeneous solutions 
$\bU^{{\rm hom},j}$ 
are not stored, we obtain the final solution by repeating steps
{\it (p,i)} and {\it (p,ii)}, but using 
the coefficients $c_j$ as inhomogeneous Dirichlet 
boundary conditions for $P$.

We point out several final aspects of this influence matrix method:
\begin{itemize}
\item As mentioned previously, the differential operators are
decoupled by azimuthal Fourier mode; this also applies to the 
influence matrix, making its size manageable, 
i.e.~$(2N_r + N_z)\times(2N_r + N_z)$ for each $m$.
\item Because the system \eqref{eq:infmat} is in fact overdetermined,
the influence matrix must be regularized (e.g.~zero eigenvalues 
replaced by finite values) in order to be inverted.
\item Although $\divv$ and $\lap$ commute, even for the discretized 
operators, this is no longer the case when boundary conditions 
are substituted for the highest order equations,
essentially because differentiation lowers the polynomial order.
The \emph{tau correction} keeps track of the derivatives of
the dropped frequencies, and, once the solution is obtained, the
necessary correction is applied to it. In the code, the tau correction is
implemented in the influence matrix, along with the boundary conditions.
\end{itemize}

\subsection{Computing details}
We used a simulation code written in Fortran in 1980s \cite{Tuc}, slightly
modernized and optimized for vector platforms
NEC SX-5 and NEC SX-8. 
The Fourier transforms used to convert from the spectral to the physical 
representation (in which the nonlinear term is computed at each timestep) 
were carried out by the NEC emulation of the SciLib Cray library.
A typical three-dimensional nonlinear run of 1000 timesteps required 
70 CPU seconds. Patterns were visualized using the VTK library.
The color map we used is based on CMRmap~\cite{CMR}, in which luminosity 
changes monotonically. Thus visualizations are correctly 
rendered in greyscale as well as in color.

\section{Results}
\subsection{Insulating sidewalls}
\label{ch:convpats:adiab}

We have performed a sequence of simulations, varying the
initial state and the Rayleigh number, in order to find the asymptotic state
for each configuration.  We chose to use Neumann boundary conditions for the
temperature deviation on the sidewalls, which corresponds to perfect
insulation.  This should be close to the experimental setup, where Perspex
plexiglass (a poor thermal conductor) was used.  

\subsubsection{Experimental and numerical protocol}

In their experiment, Hof \etal{}~\cite{HofLucMul} produced the large number of
convective patterns at $Ra=14\,200$ shown in figure~\ref{fig:hof} by
increasing and decreasing the Rayleigh number in a variety of ways.  The
three-roll state in (\textit{a}) was obtained by starting from a subcritical
Rayleigh number and gradually increasing $Ra$ in steps of $\Delta Ra=200$. A
further increase of $Ra$ to $21\,200$ followed by a decrease to $14\,200$ led
to the two-roll state in (\textit{b}).  The four-roll state~(\textit{d}) was
obtained by a sudden increase of $Ra$ from a subcritical value to $13\,000$
followed by a slow increase to $14\,200$, while the axisymmetric
state~(\textit{h}) was obtained by a sudden increase to $15\,000$ followed by
a slow decrease.  The inverted~(\textit{c,e}) and mercedes
patterns~(\textit{f,g}), as well as rotating and pulsating patterns not shown
in figure~\ref{fig:hof}, were obtained by more complicated sequences.

Rather than filling in the entire set of combinations of Rayleigh numbers and
initial conditions, we attempted to probe this parameter space.  We
initialized the simulations with a slightly perturbed conductive solution.
This can be seen as corresponding to a sudden jump of heating power in an
experimental setup, where previously a fluid was maintained below the
convection threshold.  Such simulations gave us different patterns, depending
on Rayleigh number.  Once we obtained these stable convective flows, we used
them as initial states at other Rayleigh numbers.  This is again comparable to
an experimental situation in which, once a pattern is stabilized, the heating
power is changed abruptly.

In order to qualify a solution as stable, we monitored the evolution of the
flow structure, its energy and the azimuthal velocity at two arbitrarily
chosen points. We qualified a state as stable if the observed quantities did
not change more than about $0.1 \%$ over 2--4 diffusive time units and seemed
to saturate.  We were especially careful about classifying as stable a state
we qualified already as transitional for another Rayleigh number.  We cannot,
however, exclude the possibility that a long-lasting transitional state was
interpreted as stable; see our companion paper~\cite{Boronska_PRE2}.

The field visualized throughout this article is the temperature deviation from
the basic vertical profile, referred to merely as the temperature, as for
fixed $z$ they differ only by a constant. The horizontal cuts are taken in the
midplane; dark (bright) areas indicate hot (cold) zones. The time is expressed
in dimensionless units of the vertical thermal diffusion time
$[t]=d^2/\kappa$.

\begin{figure}[!htbp]
	\centering
\includegraphics[scale=0.6]{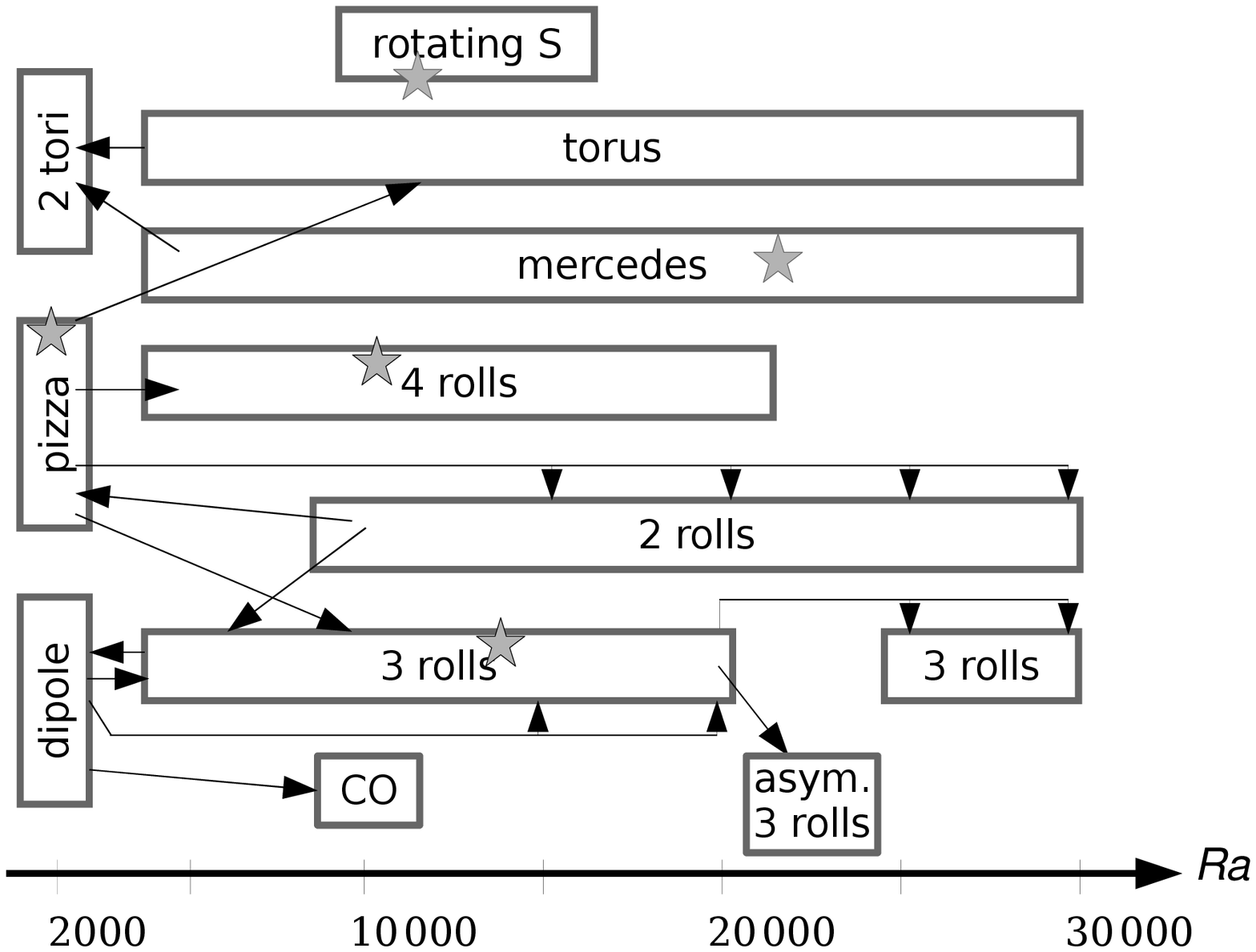}
	\caption{Schematic diagram of stability ranges and transitions between
          convective patterns as a function of Rayleigh number for $\Gamma=2$,
          $Pr=6.7$ and insulating sidewalls. Stars denote solutions obtained
          from a sudden start to the Rayleigh number indicated, using as an
          initial condition a slightly perturbed conductive state
          (see figure \ref{fig:hof:firstpats}).}
	\label{fig:hof:adiab:sumgraph}
\end{figure}

The patterns we observed and the transitions
between them are described in the sections which follow. In order to orient
the reader, we anticipate these results and present on figure
\ref{fig:hof:adiab:sumgraph} a schematic diagram of the stable patterns we
observed for different Rayleigh numbers.  More quantitative 
diagrams can be found at the end of section \ref{ch:convpats:adiab}.

\subsubsection{Sudden start from a perturbed conductive state}
\label{Evolution from perturbed conductive state}

\begin{figure}[!htbp]
\includegraphics{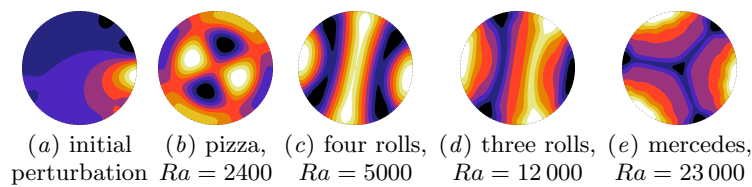}
\caption{(Color online) Initial perturbation and convective patterns obtained
from a sudden start to Rayleigh numbers indicated, using as an 
initial condition a perturbed conductive state.
  Visualization of the temperature in the cylinder's midplane,
  dark: hot/ascending, bright: cold/descending fluid.}
\label{fig:hof:firstpats}
\end{figure}
We first report the results obtained from the slightly perturbed 
conductive solution, shown in figure \ref{fig:hof:firstpats}\textit{a}.  
We run a series of simulations for
Rayleigh numbers between $1600$ and $23\,000$. Depending on the Rayleigh
number, this state evolves towards different flows shown on figure
\ref{fig:hof:firstpats}\textit{b}-\textit{e}.

For $Ra\lesssim 1900$, the initial perturbation decays to zero, i.e.~the
conductive state. For $Ra$ near $2000$, the final state is a quadrupole
pattern which we will call the \emph{pizza state} (shown
on figure \ref{fig:hof:firstpats}\textit{b}).  This state has four
well-separated sections, resembling pieces of a pizza. Each section has either
a hot round spot in the center with colder area along the sidewall, or a cold
round spot in the center with warmer area at the sidewall.

For $Ra$ between approximately $3000$ and $20\,000$, 
the system evolves towards states with parallel or rather quasi-parallel rolls.
Between $3000$ and $10\,000$ the final state has four rolls, as in the 
example depicted in figure \ref{fig:hof:firstpats}\textit{c}, which has upflow 
(light areas) along the central diameter and two small regions on the 
left and right boundaries, with downflow (dark areas) in between. 
Between $10\,000$ and $20\,000$, the final solution has three rolls;
that in figure \ref{fig:hof:firstpats}\textit{d} has upflow to the right 
and downflow to the left of the central diameter, delimiting a wide central 
roll, with a smaller roll on either side.
Finally, for $Ra \approx 23\,000$, the final pattern
consists of three radial spokes of cold descending fluid, 
named the {\em mercedes} pattern by Hof~\cite{HofThesis} (figure
\ref{fig:hof:firstpats}\textit{e}).  For all these patterns, the roll
boundaries become thinner as the Rayleigh number is increased.

In Hof's experiments, a sudden increase to $Ra=13\,000$ or $Ra=15\,000$ led to
a four-roll and an axisymmetric pattern, respectively.  However,
figure~\ref{fig:hof:adiab:sumgraph}, which summarizes the results of all of our
simulations, shows that we were able to produce four-roll and axisymmetric
patterns at these Rayleigh numbers by other routes.

\subsubsection{Three-roll patterns}
\label{sec:hof:three:rolls} 
\begin{figure}[!htbp]
\includegraphics{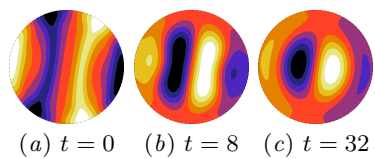}
\caption{(Color online) Evolution of the convective pattern at $Ra=2000$ for a simulation 
initialized with a three-roll pattern.}
\label{fig:hof:3roll2dip}
\end{figure}

In the second series of simulations, we used as the initial condition the
three-roll state previously converged at $Ra=14\,200$, like that in
figure \ref{fig:hof:firstpats}\textit{d}.  Below the critical
Rayleigh number this pattern decays to zero via an intermediate \emph{dipole
  pattern} and for $Ra=2000$ the three rolls transform into a dipole
state, as shown on figure~\ref{fig:hof:3roll2dip}. The dipole pattern
(\textit{c}) resembles an $m=1$ azimuthal mode.

\begin{figure}[!htbp]
\includegraphics{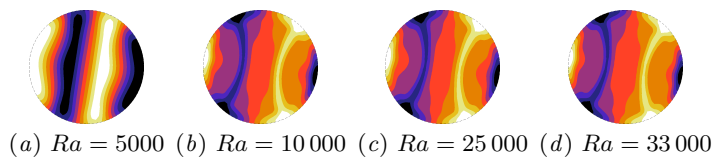}
\caption{(Color online) Three-roll patterns converged at different Rayleigh numbers.}
\label{fig:hof:3r:varra}
\end{figure}

\begin{figure}[!htbp]
\includegraphics{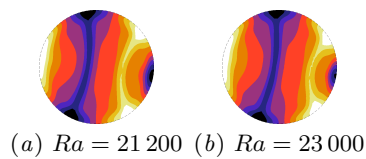}
\caption{(Color online) Asymmetric three-roll pattern with central roll shifted to the right,
  obtained for $Ra$ between $20\,000$ and $25\,000$.}
\label{fig:hof:imp3r}
\end{figure}

\begin{figure}[!htbp]
\includegraphics{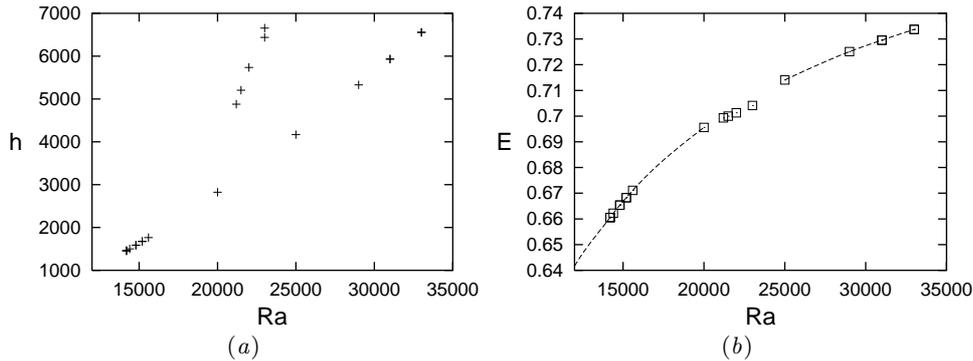}
\caption{Dependence of the three-roll patterns on Rayleigh number:
(\textit{a}) temperature at $r=0.3$, $z=0.25$, $\theta=0$; (\textit{b}) energy.}
\label{fig:hof3stat}
\end{figure}

For $Ra=5000$ and above, up to $Ra=33\,000$, the three-roll state remains
stable, with the rolls more curved for higher Rayleigh numbers (see figure
\ref{fig:hof:3r:varra}).  An exception is the range between $Ra=20\,000$ and
$Ra=25\,000$, where this pattern becomes asymmetric -- the band of colder fluid between the central roll and the
left roll moves slightly towards the center as the Rayleigh number is
increased (see figure~\ref{fig:hof:imp3r}).  This is similar to the results of
Hof \etal{}~\cite{HofLucMul}, although they found that
the leftmost roll vanishes eventually at $Ra=21\,000$, where the flow forms a
two-roll pattern. 
Figure \ref{fig:hof3stat} shows the Rayleigh-number dependence of 
the temperature at a fixed point and of the energy;
both show deviations in the range $20\,000<Ra<25\,000$.
We conclude that a bifurcation is present.  In the experiment this
evolution yielded a two-roll pattern and, during our further analysis of
two-roll states, we found that their energy indeed approaches that of shifted
(asymmetric) three rolls.

\subsubsection{Evolution from four rolls}
\begin{figure}[!htbp]
\includegraphics{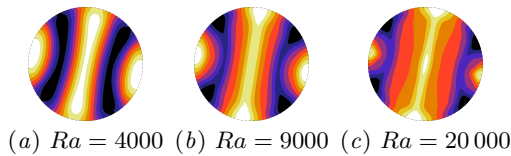}
\caption{(Color online) Four-roll patterns for different Rayleigh numbers.}
\label{fig:4r:varra}
\end{figure}

\begin{figure}[!htbp]
\includegraphics{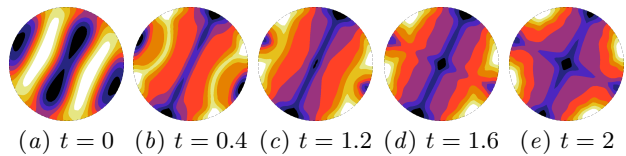}
\caption{(Color online) Evolution of the four-roll pattern towards (probably transitional) cross pattern at $Ra=25\,000$.}
\label{fig:hoft4r}
\end{figure}

We used a four-roll pattern converged at $Ra=3000$, like that in figure 
\ref{fig:hof:firstpats}\textit{c} as an initial condition for
simulations at $4000 \leq Ra \leq 20\,000$.  The newly evolved patterns are
also of the four-roll family, and, as in the case of three rolls, 
the roll boundaries become thinner as the Rayleigh number is increased 
(see figure \ref{fig:4r:varra}).
We then used the four-roll state converged for $Ra=20\,000$ 
depicted in figure~\ref{fig:4r:varra}\textit{c} as an initial
condition at Rayleigh numbers $25\,000$ and $29\,000$. This time the geometry
of the pattern changes, as displayed on figure~\ref{fig:hoft4r} -- the
four-roll pattern turns into a \emph{cross pattern} with four spokes of
descending cold fluid.  The cross flow did not saturate in the time 
during which we observed it, but continued to evolve slowly; 
we suspect that it is a transitional rather than an asymptotic state.
This would be in agreement with Hof~\cite{HofPriv}, who observed that the cross
pattern is a long-lasting transient state unstable to a mercedes pattern.

\subsubsection{Evolution from pizza pattern}
\label{hofadiab:Evolution from pizza pattern}
\begin{figure}[!htbp]
	\includegraphics{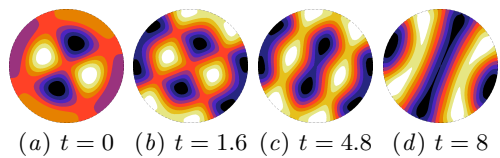}
	\caption{(Color online) Evolution from pizza pattern at $Ra=5000$.}
	\label{fig:hofmp:evol}
\end{figure}

\begin{figure}[!htbp]
\includegraphics{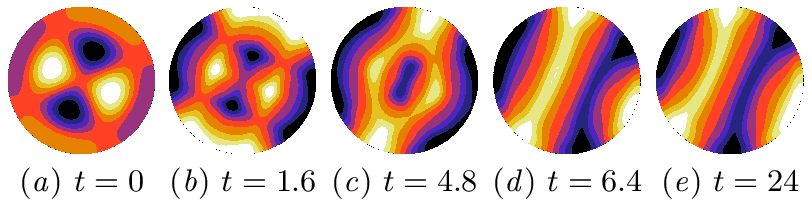}
\begin{center}
\caption{(Color online) Evolution from pizza pattern at $Ra=10\,000$.}
	\label{fig:hofnp:evol}
\end{center}
\end{figure}

\begin{figure}[!htbp]
\includegraphics{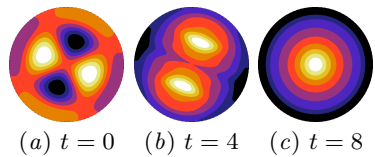}
\caption{(Color online) Evolution of pattern at $Ra=14\,200$: from initial pizza  through eight towards final torus pattern.}
\label{fig:hofqp:evol}
\end{figure}

We used the pizza pattern at $Ra=2000$, like that in figure 
\ref{fig:hof:firstpats}\textit{b}, as an initial condition for a
series of simulations at several Rayleigh numbers between $5000$ and
$29\,000$.  In this range the pizza pattern is not stable.  For $Ra=5000$ it
changes into four rolls (see figure~\ref{fig:hofmp:evol}) and for
$Ra=10\,000$ into three rolls (figure~\ref{fig:hofnp:evol}).  For $Ra=14\,200$
the initial pizza flow evolves into a \emph{torus pattern} -- an axisymmetric
state with one toroidal roll, passing through an intermediate
\emph{eight pattern}.  This evolution is displayed on figure
\ref{fig:hofqp:evol}.  The transitional eight pattern was also observed by
Hof~\cite{HofPriv}.

\begin{figure}[!htbp]
\includegraphics{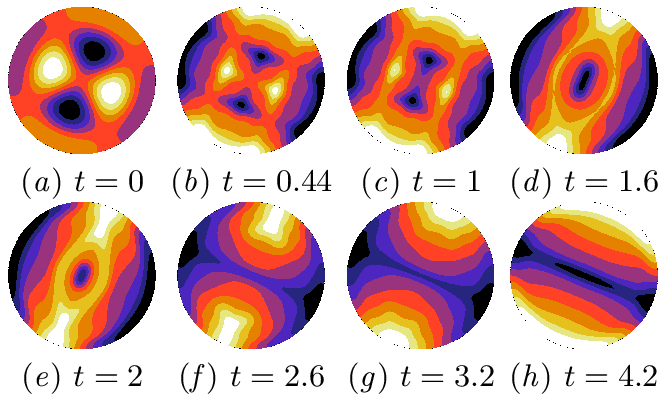}
	\caption{(Color online) Evolution of pattern at   $Ra=16\,000$ from initial pizza through a series of intermediate states towards the final two-roll pattern.} 
	\label{fig:hofrp:abc}
\end{figure}

\begin{figure}[!htbp]
\includegraphics{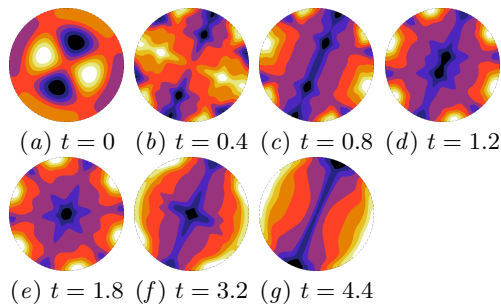}
\caption{(Color online) Evolution of pattern at $Ra=29\,000$ from initial pizza through a series of intermediate states towards the final two-roll pattern.}
	\label{fig:hofup}
\end{figure}

For $Ra \geq 15\,000$ the pizza state, after passing through a series of
various transitional patterns, eventually evolves to a two-roll flow. Figure
\ref{fig:hofrp:abc} displays the evolution of the system for $Ra=16\,000$,
where we describe the intermediate states as: \emph{triangle mosaic}
(\textit{b}), \emph{eye} (\textit{d}) and \emph{imperfect eight}
(\textit{f}). Figure~\ref{fig:hofup} presents the system behaviour for
$Ra=29\,000$, where the intermediate patterns between pizza and two rolls are
\emph{six-spoke elongated-star} (\textit{d}) and \emph{six-spoke star}
(\textit{e}).


\subsubsection{Evolution from two rolls}
\begin{figure}[!htbp]
\includegraphics{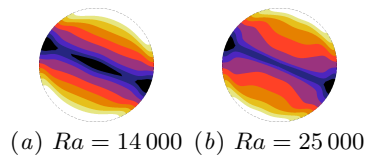}
\caption{(Color online) Two-roll pattern at different Rayleigh numbers.}
\label{fig:hof2r:varra}
\end{figure}

Another initial condition we used was the two-roll state
like that in figure \ref{fig:hofrp:abc}\textit{h}, converged previously
at $Ra=15\,000$.  For $Ra=2000$ it leads to the pizza pattern, and for
$Ra=5000$ a three-roll state.  For $10\,000\leq Ra < 29\,000$, we found the
two roll pattern to be stable. At $Ra=29\,000$ the roll boundaries start oscillating
slightly.  Figure~\ref{fig:hof2r:varra} presents two-roll flow visualizations
for different Rayleigh numbers.

\subsubsection{Axisymmetric flows}
\begin{figure}[!htbp]
\begin{center}
\includegraphics{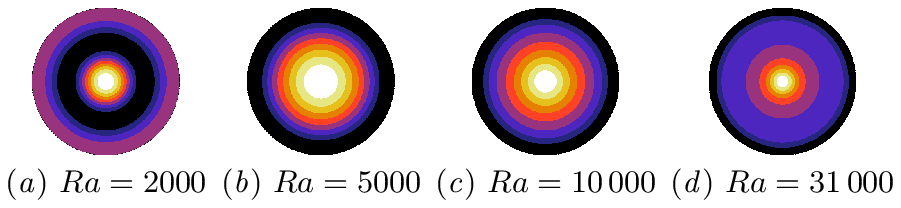}
\end{center}
	\caption{(Color online) Axisymmetric patterns at different Rayleigh numbers.}
	\label{fig:hof:axi:varra}
\end{figure}

The axisymmetric pattern, shown in figure \ref{fig:hofqp:evol}\textit{c} 
and used as an initial
condition, also leads to axisymmetric patterns for a wide range of Rayleigh
numbers $2000 \leq Ra \leq 33\,000$ (see figure~\ref{fig:hof:axi:varra}).  For
$Ra=2000$ there are two concentric toroidal rolls instead of one.  This is in
partial agreement with Hof~\cite{HofThesis}, who found toroidal flow 
to be stable for $Ra > 3500$, but to evolve towards a rotating three-petaled
pattern for $Ra\gtrsim 23\,000$.  
At these values of $Ra$, our simulations still converge towards a 
stationary axisymmetric flow, although in an oscillatory manner.  
All axisymmetric patterns we observed were truly
two-dimensional, with no azimuthal velocity.
	
\subsubsection{Evolution from mercedes pattern}
\begin{figure}[!htbp]
\includegraphics{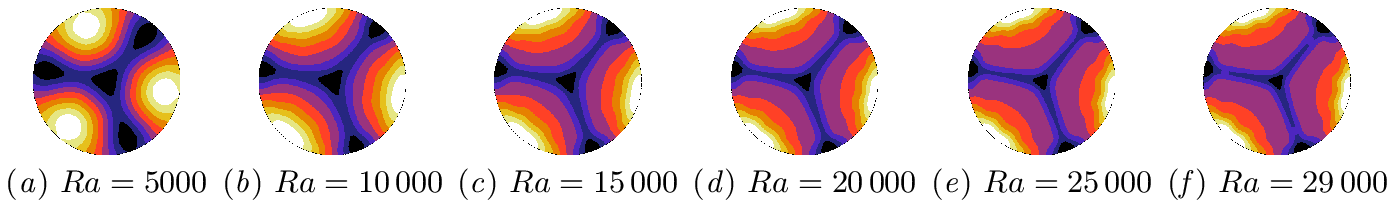}
	\caption{(Color online) Mercedes patterns at different Rayleigh numbers.}
	\label{fig:hof:merc:varra}
\end{figure}

We ran a series of simulations using as the initial condition the mercedes
pattern of figure \ref{fig:hof:firstpats}\textit{e}.
For every Rayleigh number in the range $5000
\leq Ra \leq 29\,000$ this gave stable mercedes patterns and for $Ra=2000$ it
evolves to a two-roll axisymmetric pattern.  Figure~\ref{fig:hof:merc:varra}
displays the final patterns for different Rayleigh numbers.

\subsubsection{Evolution from dipole pattern}
\begin{figure}[!htbp]
\includegraphics{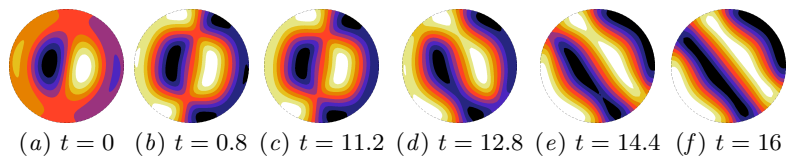}
	\caption{(Color online) Evolution from dipole pattern at $Ra=5000$.}
	\label{fig:hofmD:evol}
\end{figure}
\begin{figure}[!htbp]
\includegraphics{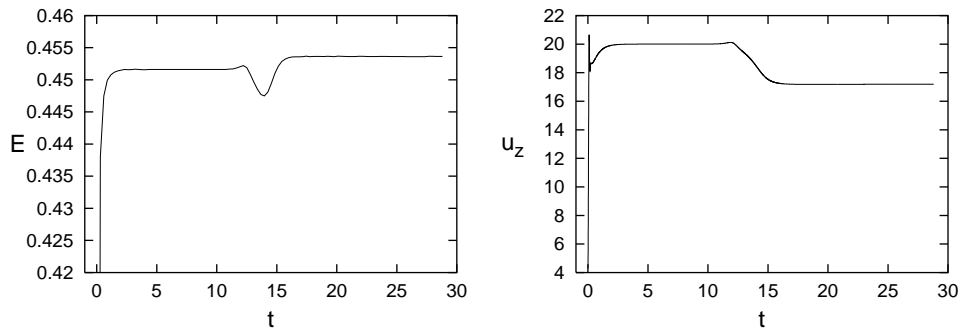}
	\caption{Evolution from the dipole pattern at $Ra=5000$: (\textit{a}) energy, (\textit{b}) vertical velocity at one point. A long-lasting transient state is visible between $t=3$ and $t=12$.}
	\label{fig:hofmD:uen}
\end{figure}
\begin{figure}[!htbp]
	\centering
	\includegraphics{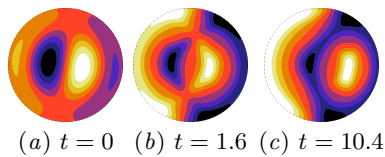}
	\caption{(Color online) Evolution from dipole pattern through dipole smile into final CO pattern at $Ra=10\,000$.}
	\label{fig:hofnD}
\end{figure}
\begin{figure}[!htbp]
	\centering
	\includegraphics{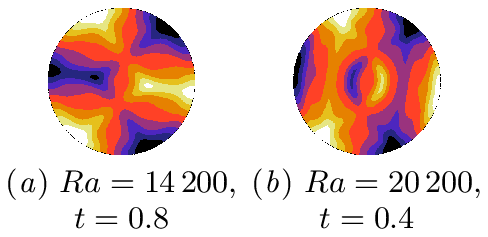}
	\caption{(Color online) Transitional patterns observed during evolution from dipole into three-roll pattern.}
	\label{fig:hof:D:tranpats}
\end{figure}
\begin{figure}[!htbp]
	\centering
		\includegraphics{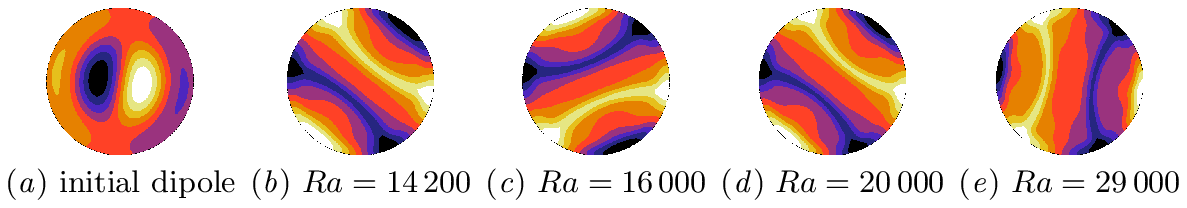}	
                \caption{(Color online) Initial dipole (a) and final patterns for various Rayleigh
                  numbers.}
	\label{fig:hof:D:finpats}
\end{figure}

\begin{figure}[!htbp]
	\centering
		\includegraphics{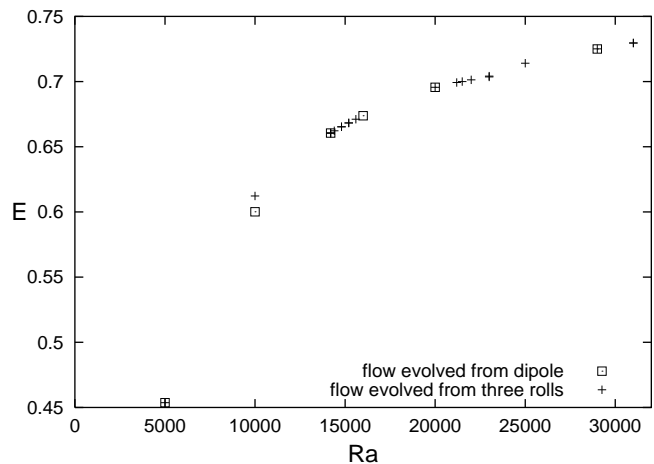}
	\caption{Squares: energy of the flow evolved from dipole state as a function of Rayleigh number; crosses: energy of the flow evolved from three-roll state (for comparison). At $Ra=10\,000$ the dipole pattern evolves into a CO pattern.}
	\label{fig:hof:D:en}
\end{figure}

Another initial condition we used was the dipole pattern 
shown in figure \ref{fig:hof:3roll2dip}\textit{c}.
For Rayleigh numbers between $5000$ and $29\,000$, with
the exception of $10\,000$, the flow evolves towards a three-roll pattern.
The evolution of the flow at $Ra=5000$ is presented on figures
\ref{fig:hofmD:evol} and \ref{fig:hofmD:uen}. At this Rayleigh number the
system passes through a long-lasting intermediate \emph{dipole smile state}
(figure~\ref{fig:hofmD:evol}\textit{c}), whose lifetime is of order $10$.
The flow patterns appearing for $Ra=10\,000$ are depicted on figure
\ref{fig:hofnD}.  An intermediate dipole smile state appears also
(\ref{fig:hofnD}\textit{b}), with a lifetime of about $2$.  The final solution
is a \emph{CO pattern} (\ref{fig:hofnD}\textit{c}), composed of one curved and
one circular roll.  This state resembles the three-roll pattern with the ends
of two neighbouring rolls joined together, but its energy is higher than that
of the three-roll state at the same $Ra$.
At higher Rayleigh numbers, where the asymptotic solution is again three-roll
flow, transitional patterns also appear (figure~\ref{fig:hof:D:tranpats}). The
final patterns are shown on figure~\ref{fig:hof:D:finpats}, with the initial
dipole pattern for comparison. The initial pattern orientation does not seem
to determine the direction of the rolls of the final pattern.
Figure~\ref{fig:hof:D:en} depicts the energy of patterns evolved starting from
a dipole pattern as a function of Rayleigh number. For comparison, the energy
of three-roll states from section~\ref{sec:hof:three:rolls} is also displayed.
They are in perfect agreement with the sole exception of $Ra=10\,000$: the
energy of the CO state is lower.

\subsubsection{Dipole-shaped perturbation}

As in section \ref{Evolution from perturbed conductive state},
we again used as an initial condition a low-amplitude state.
This initial condition was obtained as follows.
We began with the three-roll state converged at $Ra=14\,200$, and lowered $Ra$ 
to 1200, i.e. below the convective threshold. 
A transient dipole appeared, similar to that depicted in
figure \ref{fig:hof:3roll2dip}\textit{c}, and we halted the evolution 
before the flow reverted to the conductive state.
Although this initial condition is not a converged pattern, it 
it occurs during the evolution of the system, and is thus
reachable experimentally.

\begin{figure}[!htbp]
	\centering
	\includegraphics{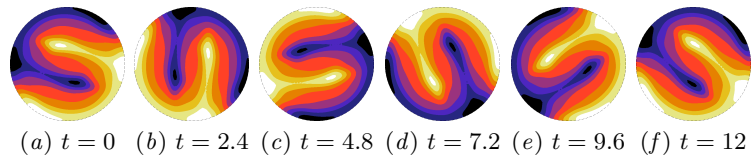}
	\caption{(Color online) Rotating S pattern at $Ra=12\,500$ at six different times.}
	\label{fig:hof:rotzip}
\end{figure}

\begin{figure}[!htbp]
	\centering
		\includegraphics{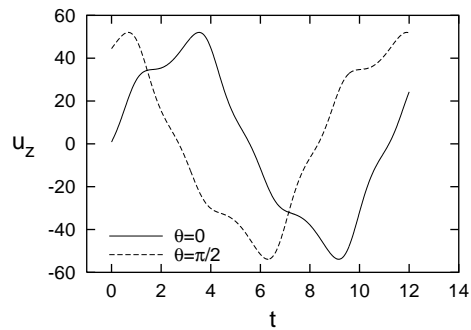}
	\caption{Evolution of vertical velocity in time for rotating S pattern at two points of the same $r$ and $z$ and different $\theta$.}
	\label{fig:hofja:evol}
\end{figure}

\begin{figure}[!htbp]
\includegraphics{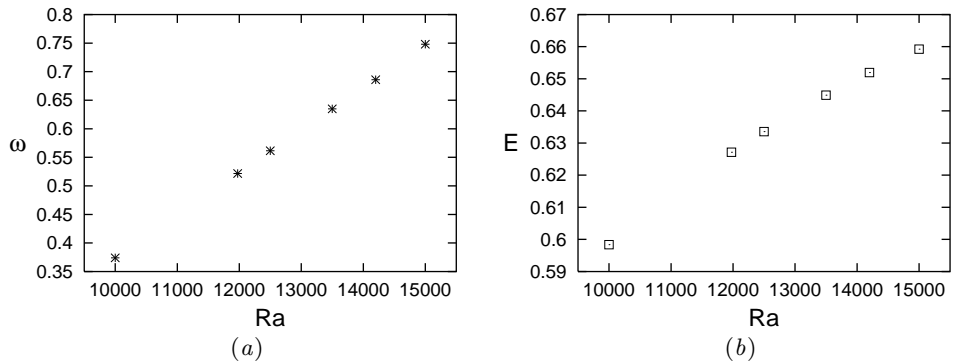}
	\caption{Frequency and energy of the rotating S pattern as a function of Rayleigh number.}
	\label{fig:hof:rotzip:enfr}
\end{figure}

Starting with this initial condition, simulation yields a dipole at $Ra=2000$
and three rolls at $Ra=5000$.  However, for $10\,000 \leq Ra \leq 15\,000$, we
obtain a new state.  A transitional dipole smile gives way to a
slowly rotating roll in the shape of the letter S, which we will refer to as a
\emph{rotating S pattern}.  
This time-dependent state resembles a truncated version of the rotating spiral 
found in simulations in a $\Gamma=4$ geometry~\cite{RudFeu}.
A visualization of the rotating S at 
different times is displayed on figure~\ref{fig:hof:rotzip} and the evolution
of the temperature at two points is plotted in figure~\ref{fig:hofja:evol}.
The rotation is very slow: one period is of the order of ten, and the
frequency grows with the Rayleigh number.  Figure~\ref{fig:hof:rotzip:enfr}
shows the dependence of the frequency and energy of the rotating S on
Rayleigh number.

\begin{figure}[!htbp]
	\centering
 	\includegraphics{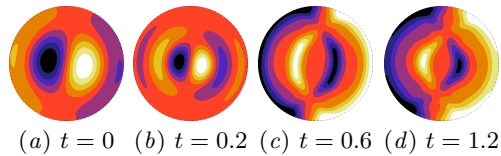}
	\caption{(Color online) Evolution from dipole-shaped perturbation into a stable dipole smile pattern at $Ra=16\,000$.}
	\label{fig:hofrda}
\end{figure}

The evolution of the dipole pattern at $Ra=16\,000$ is shown on figure 
\ref{fig:hofrda}. It passes through an intermediate \emph{three-part dipole pattern} before finally becoming a dipole-smile pattern. This pattern, observed for 
$5000\leq Ra \leq 15\,000$  as a transient pattern, seems to be stable for this Rayleigh number.

For $20\,000 \leq Ra \leq 29\,000$, the dipole pattern transforms at first
into a transitional pattern, similar to dipole smile, then into an S roll and
finally it becomes a stable three-roll flow.  The transition occurs without
any oscillatory evolution and the energy of the final state fits the previously
observed dependence between three-roll pattern energy and Rayleigh number of 
figure \ref{fig:hof:D:en}.

\subsubsection{Summary diagrams}
\label{Adiab:Sumdiag}
\begin{figure}[!htbp]
\centering
	\includegraphics[width=14cm]{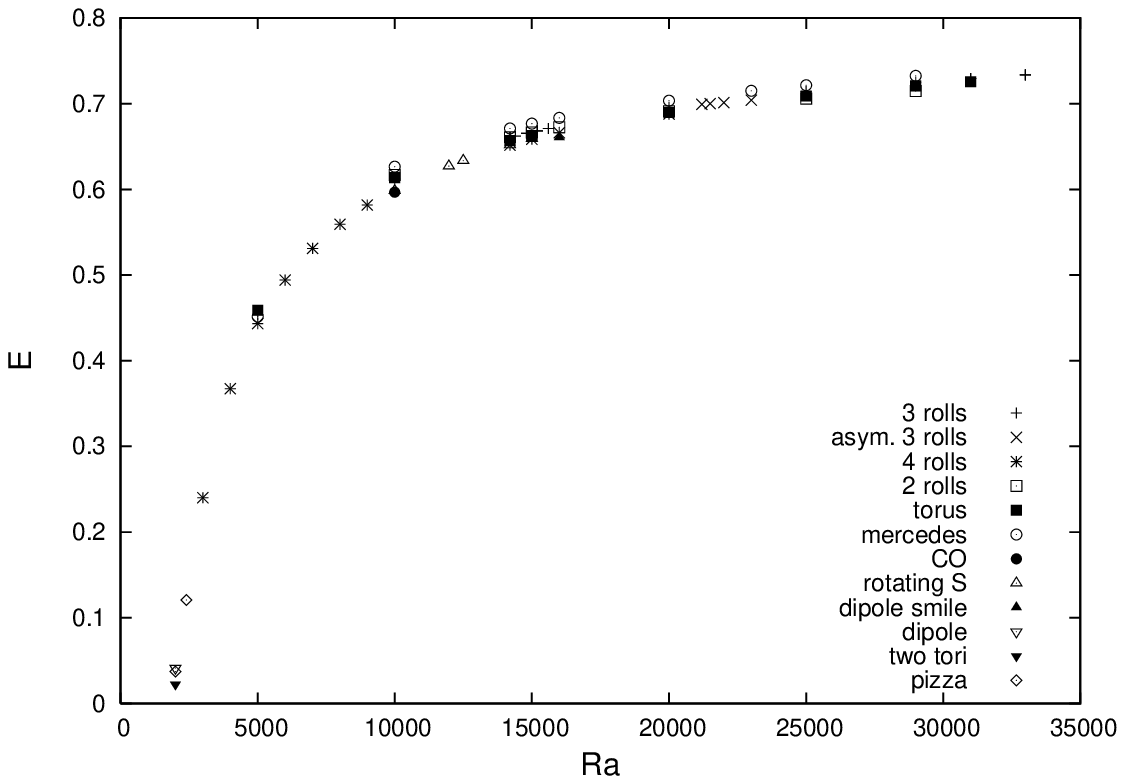}
	\caption{Energy as a function of Rayleigh number 
of all stable convective patterns obtained for insulating sidewalls.}
	\label{fig:hof_adiab_endiag}
\end{figure}

The energy of all stable patterns described above, 
as a function of Rayleigh number, 
is depicted on figure~\ref{fig:hof_adiab_endiag}.
\begin{figure}[!htbp]
	\centering
		\includegraphics[width=14cm]{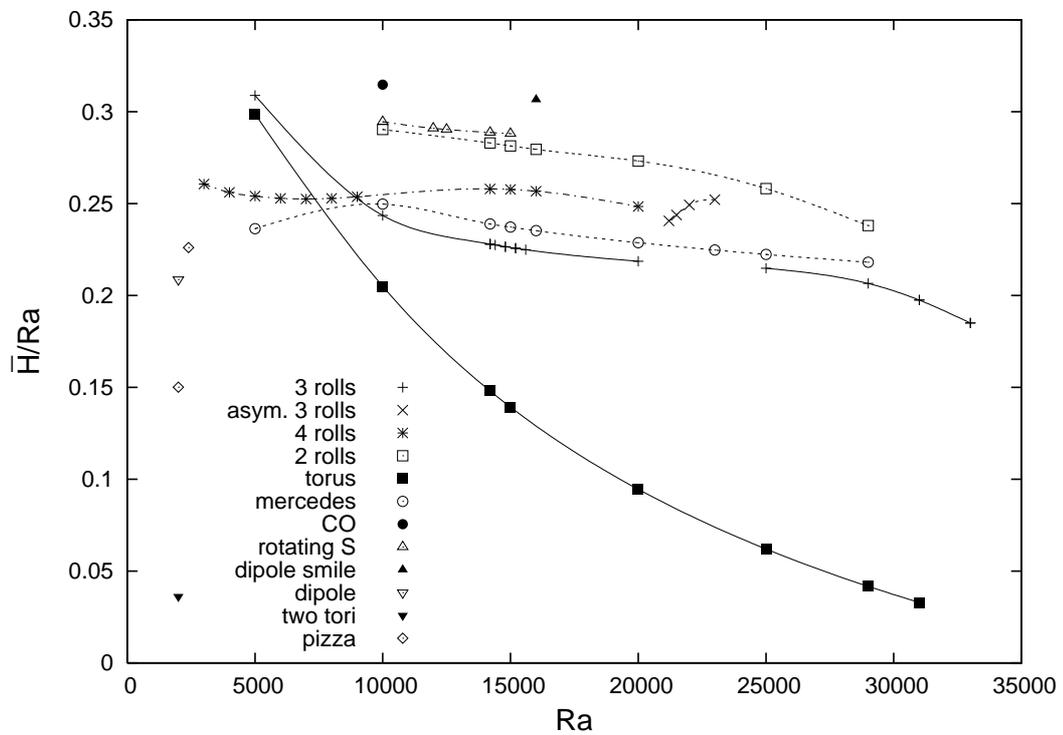}
	\caption{Scaled temperature deviation $\Hmax/Ra$ as a function of 
Rayleigh number for all stable convective patterns obtained for 
insulating sidewalls.}
	\label{fig:hof:adiab:bifdiag}
\end{figure}
The energy depends on the type of convective pattern, as well as on the 
$Ra$, but the values for different patterns obtained at the same $Ra$
are very close.  


Figure \ref{fig:hof:adiab:bifdiag} shows a preliminary bifurcation diagram. 
We define $\Hmax$ to be the maximum absolute value of the temperature
deviation over the ring at $(r=0.3,\theta,z=0)$
\begin{equation}
\Hmax \equiv \max_\theta | H(r=0.3,\theta,z=0) |.
\label{eq:defh}\end{equation}
$\Hmax$ itself (as well as the commonly used Nusselt number)
has a strong linear dependence on $Ra$; plotting them directly
as a function of $Ra$ does little to separate the branches 
and this is why we have chosen to represent each state by 
its value of $\Hmax/Ra$. 



\subsection{Conducting sidewalls}
\label{ch:convpats:cond}

\begin{figure}[!htbp]
	\centering
\includegraphics[scale=0.6]{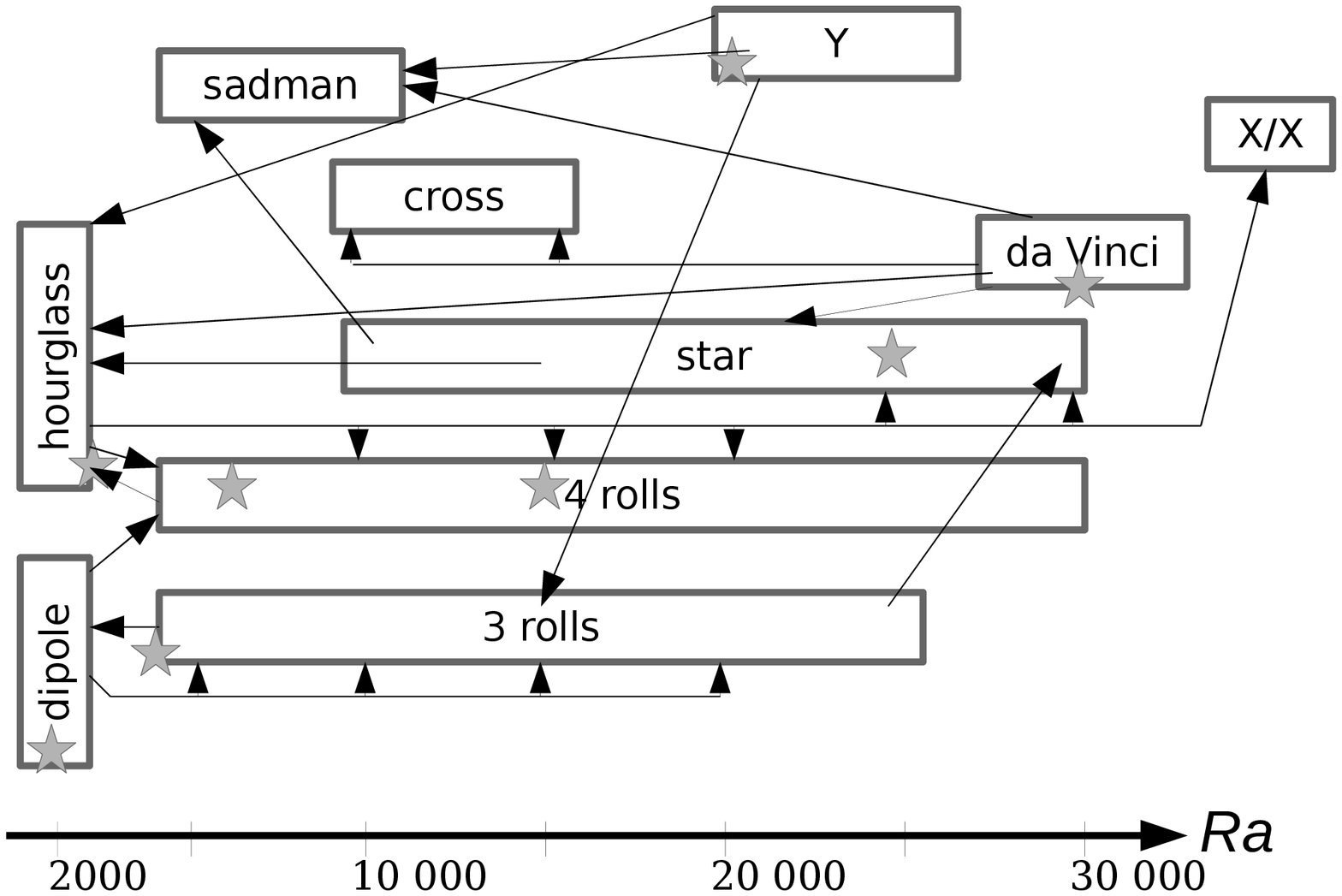}
	\caption{Schematic diagram of stability ranges and transitions between
          convective patterns as a function of Rayleigh number for $\Gamma=2$,
          $Pr=6.7$ and conducting sidewalls. Stars denote solutions obtained
          from a slight perturbation of the conductive state, at given $Ra$.
          (see figure \ref{fig:hc:firstpats}).}
	\label{fig:hof:cond:sumgraph}
\end{figure}

We now present results obtained for perfectly conducting sidewalls,
i.e. we apply homogeneous Dirichlet boundary conditions \eqref{eq:bcdir} 
instead of Neumann boundary conditions \eqref{eq:bcneu}. 
The configuration is otherwise identical, i.e.~$\Gamma=2$ and $Pr=6.7$. 
A schematic diagram organising all the results we will present is shown on
figure~\ref{fig:hof:cond:sumgraph}.  

\subsubsection{Sudden start from perturbed conductive state}

\begin{figure}[!htbp]
\centering
\includegraphics{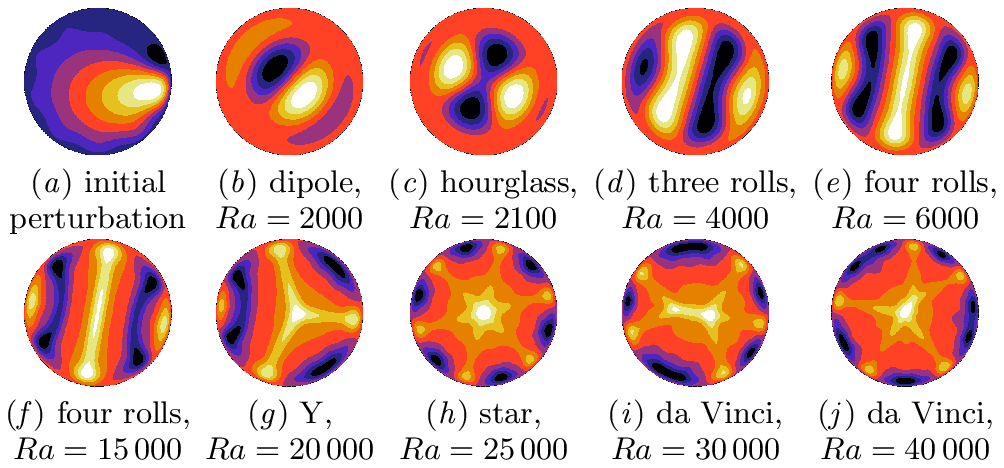}
\caption{(Color online) Arbitrary perturbation used for initialising simulations (\textit{a}) and final patterns at various Rayleigh numbers.}
\label{fig:hc:firstpats}
\end{figure}

As before, we initialized the first
series of simulations with a perturbed conductive solution at various Rayleigh
numbers between $1900$ and $40\,000$.  The initial perturbation and final
states are represented symbolically on figure~\ref{fig:hof:cond:sumgraph} and
displayed in figure~\ref{fig:hc:firstpats}. For $Ra=1900$ and
$2000$ the final state is of dipole form. For $2100\leq Ra \leq 2500$, instead
of the pizza pattern observed for the insulating case, we found an
\emph{hourglass} pattern, with elongated
cold spots touching at the center.  For $Ra=2700$ and $4000$ we
obtained three rolls and for $6000\leq Ra \leq 15\,000$ four rolls, which
differ from the analogous patterns described previously only at
the sidewalls, since, here, the deviation from the conductive profile must be zero at the
boundaries.  For $Ra=20\,000$ we observed a \emph{Y pattern} (figure
\ref{fig:hc:firstpats}\textit{g}) with three bands of hot fluid in the
shape of the letter Y. It is similar to the previously observed mercedes
pattern, but has only one and not three symmetry axes.  For $Ra=25\,000$ we
obtained a state in the form of a six-armed star, presented on figure
\ref{fig:hc:firstpats}\textit{h}).  For $Ra=30\,000$ and above, up to
$40\,000$, we obtained a pattern we call \emph{da Vinci}, because of its resemblance
with the artist's sketch of human body proportions (figures
\ref{fig:hc:firstpats}\textit{i},\textit{j}).  All of these flows were
stationary.
 
\subsubsection{Three rolls}
\label{Three rolls}
In this series of simulations we used as initial condition the three-roll
pattern of figure \ref{fig:hc:firstpats}\textit{d}. 
We then obtained stable three-roll
patterns for a wide range of Rayleigh numbers $3000 \leq Ra \leq 24\,000$ 
and $26\,000 \leq Ra \leq 29\,000$.
At $Ra=25\,000$ 
the flow seems to evolve instead towards a two-roll state. 
(We can, however, obtain a three-roll pattern at $Ra=25\,000$
by initializing the simulation with three rolls
obtained at $Ra=20\,000$.)
For $Ra=30\,000$, the simulation evolves to a star pattern.

\begin{figure}[!htbp]
\includegraphics{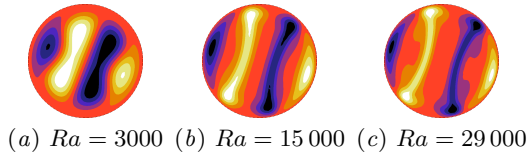}
\caption{(Color online) Stable three-roll patterns at different Rayleigh numbers.}
\label{fig:hoc:3r}
\end{figure}


\subsubsection{Four rolls}
\begin{figure}[!htbp]
\includegraphics{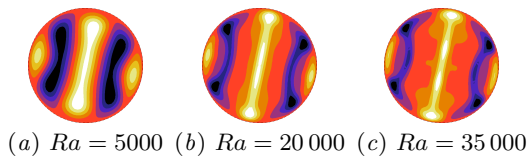}
\caption{(Color online) Stable four-roll patterns at different Rayleigh numbers.}
\label{fig:hoc:4r}
\end{figure}

Every four-roll pattern we used as initial condition retained its structure
within a large range of Rayleigh numbers $5000 \leq Ra \leq 35\,000$ (see
figure~\ref{fig:hoc:4r}).  For $Ra=2000$ the initial four-roll flow evolves
into an hourglass pattern (like that on
figure~\ref{fig:hc:firstpats}\textit{c}) and at $Ra=40\,000$ it remains a
four-roll pattern, but with the roll boundaries vibrating slightly with an
oscillation period $T=0.026$.

\subsubsection{Evolution from dipole pattern}
When we used the dipole pattern of 
figure \ref{fig:hc:firstpats}\textit{b} as an initial
condition at higher Rayleigh numbers, the flow evolved into three rolls for
$Ra=5000$ and four rolls for $10\,000\leq Ra \leq 20\,000$.  For $Ra=25\,000$
we obtained a Y pattern like that of figure
\ref{fig:hc:firstpats}\textit{g}. For $Ra=30\,000$ the simulation 
evolves initially towards a three-roll state, and so eventually should
reach a star pattern, as ascertained in section~\ref{Three rolls}.

\subsubsection{Evolution from hourglass pattern}
\begin{figure}[!htbp]
\includegraphics{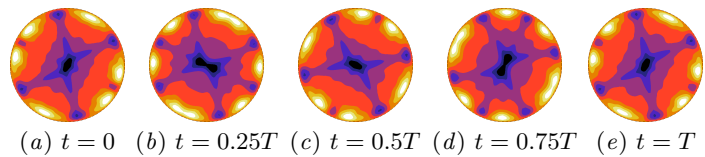}
\caption{(Color online) Oscillatory X/X pattern at $Ra=35\,000$.}
\label{fig:hoc:osc35k}
\end{figure}

For simulations initialized with the hourglass pattern shown in 
figure~\ref{fig:hc:firstpats}\textit{c} we obtained 
a stable hourglass pattern for $Ra=2000$, four-roll
flow for $5000 \leq Ra \leq 20\,000$, and a six-armed star for $Ra=25\,000$
and $30\,000$. For $Ra\geq 35\,000$ a new standing-wave pattern appears.
This state oscillates between a left-tilted and right-tilted X-letter 
shape, passing via intermediate star-like patterns; 
see figure~\ref{fig:hoc:osc35k}. The oscillation period of this 
X/X state is $T=3.05$.

\subsubsection{Evolution from star pattern}
\begin{figure}[!htbp]
\includegraphics{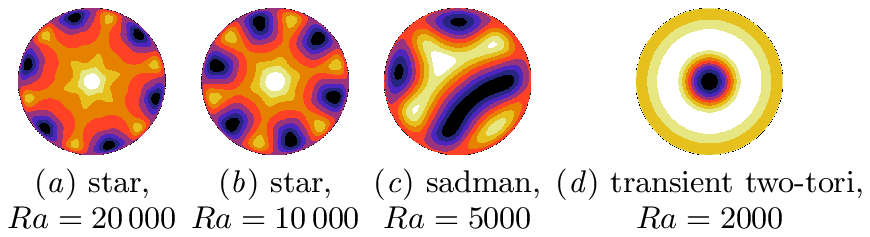}
\caption{(Color online) Convective patterns evolved from star flow.}
\label{fig:hoc:starpats}
\end{figure}

\begin{figure}[!htbp]
	\centering
		\includegraphics{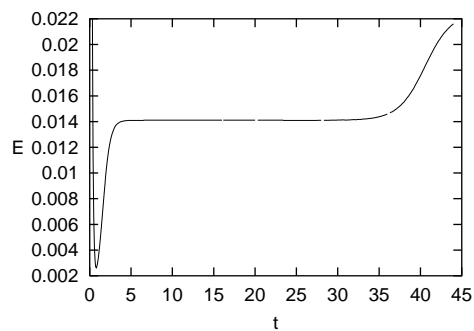}
	\caption{Evolution of the system initialized with star pattern at $Ra=2000$: energy as a function of time. Between $t=5$ and $t=35$ a long-lasting transient axisymmetric pattern exists.}
	\label{fig:hc_axi_evol}
\end{figure}

In order to obtain patterns in the form of a six-armed star, we used as
initial condition the star flow converged at $Ra=25\,000$.  Several final
convective structures obtained for different Rayleigh numbers are presented on
figure~\ref{fig:hoc:starpats}. The star flow remains stable for $10\,000 \leq
Ra \leq 30\,000$ (figure
\ref{fig:hoc:starpats}\textit{a}-\textit{b}). Below this range the initial
pattern evolved into a \emph{sadman pattern}
(figure~\ref{fig:hoc:starpats}\textit{c}) at $Ra=5000$ and an hourglass
pattern at $Ra=2000$.  For both $Ra=2000$ and $Ra=5000$, axisymmetric
transient patterns appear.  For $Ra=2000$ the transient state is composed of
two concentric toroidal rolls, and its lifetime is about $30$. Figure
\ref{fig:hc_axi_evol} shows the evolution of the energy during the transition
from the initial star flow, through a long-lasting axisymmetric state into the
asymptotic hourglass pattern.  While the flow is axisymmetric,
the energy remains almost constant.
	
\subsubsection{Evolution from da Vinci pattern}
\begin{figure}[!htbp]
\begin{center}
\includegraphics[width=1.5cm]{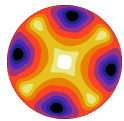} 
\end{center}
\caption{(Color online) Cross pattern evolved at $Ra=10\,000$ from da Vinci flow.}
\label{fig:hoc:cross}
\end{figure}

The da Vinci pattern (figure~\ref{fig:hc:firstpats}\textit{i}), used as
initial condition, was stable only for $Ra\geq30\,000$.  
For lower Rayleigh numbers the simulation evolves towards 
hourglass at 2000, sadman at 5000, a
cross at $14\,200 $ and $10\,000$ (figure~\ref{fig:hoc:cross}), 
and a stable five-armed star at $20\,000$ and seemingly also at $25\,000$.

\subsubsection{Evolution from Y pattern}
Having obtained the Y pattern at $Ra=20\,000$ we reused it as an initial
condition.  We found stable Y flows for Rayleigh numbers $20\,000 \leq Ra 
\leq 25\,000$.
For $Ra=14\,200$ the evolution led to three rolls.  For lower Rayleigh numbers
we obtained sadman patterns at $Ra=10\,000$ and $Ra=5000$ and hourglass at
$Ra=2000$.  Since there is a resemblance between the sadman and Y pattern,
these may be related.




\subsubsection{Summary diagrams}
\begin{figure}[!htbp]
\centering
	\includegraphics[width=14cm]{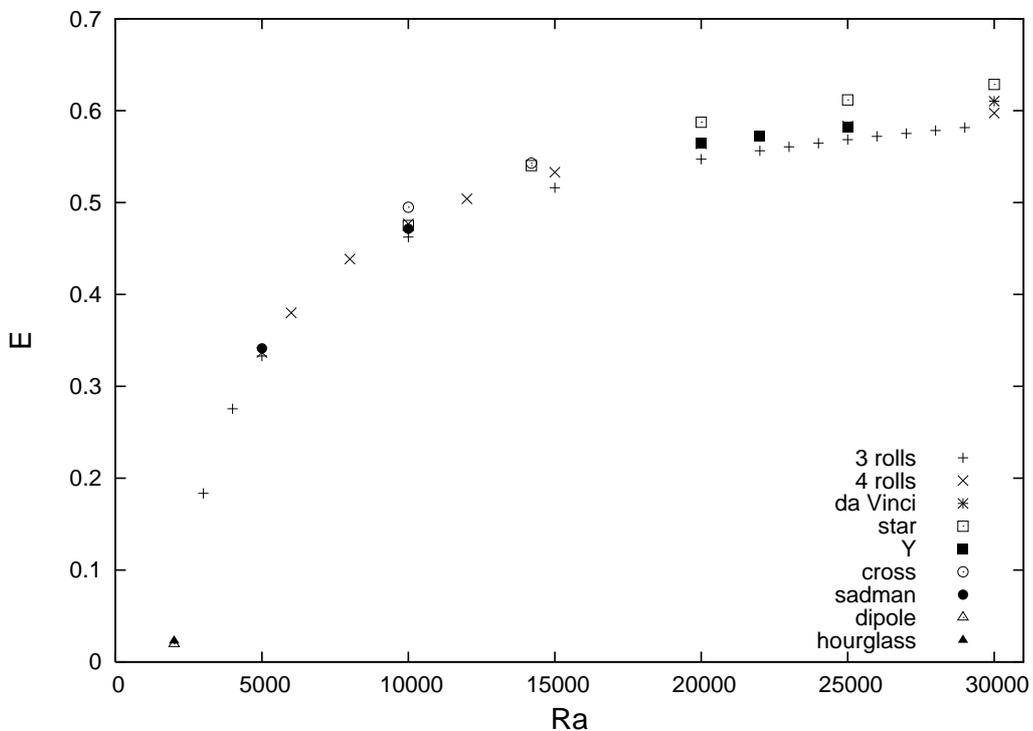}
	\caption{Energy as a function of Rayleigh number 
of all stable convective patterns obtained for conducting sidewalls.}
	\label{fig:hof_cond_endiag}
\end{figure}

\begin{figure}[!htbp]
	\centering
		\includegraphics[width=14cm]{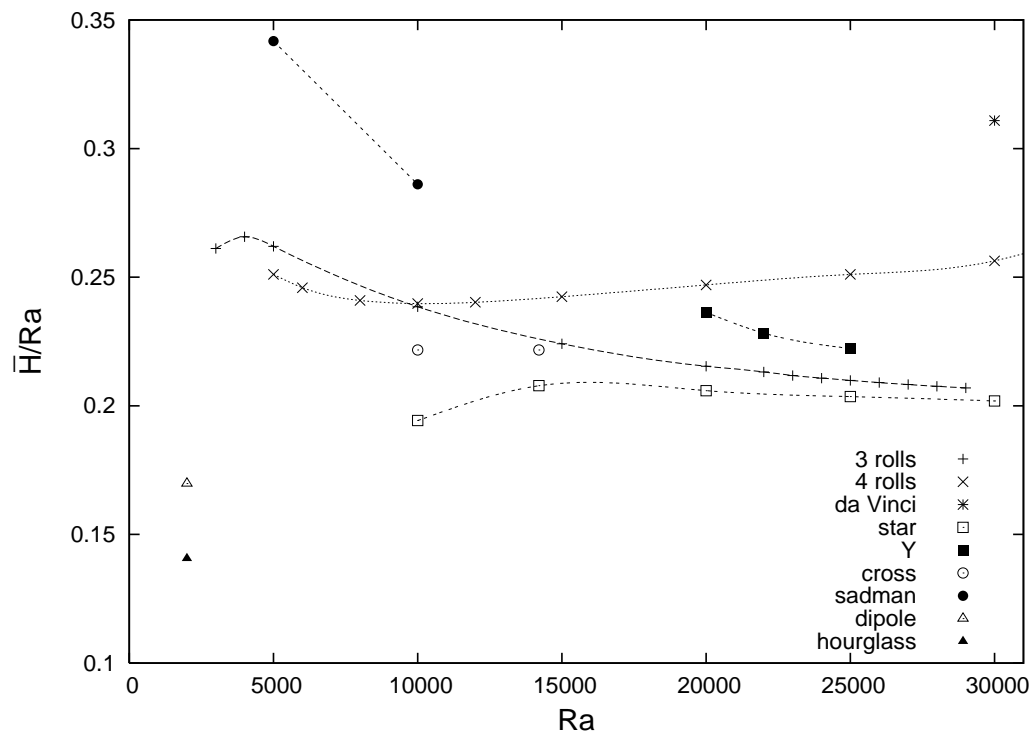}
	\caption{Scaled temperature deviation $\Hmax/Ra$ as a function of 
Rayleigh number for all stable convective patterns obtained for 
conducting sidewalls.}
	\label{fig:hof:cond:bifdiag}
\end{figure}

Figure~\ref{fig:hof_cond_endiag} shows the energy of all stable patterns found
for Dirichlet thermal boundary condition.  As in the case of insulating
sidewalls, the energy depends primarily on Rayleigh number, with a slight
variation between types of convective patterns.
Figure~\ref{fig:hof:cond:bifdiag} shows $\Hmax/Ra$, as defined in \eqref{eq:defh}, 
 plotted as a
function of Rayleigh number, for all stable patterns found.  For higher
Rayleigh numbers, several patterns remain stable over large intervals of $Ra$:
three rolls, four rolls and star.  Patterns da Vinci, Y and cross were
observed in smaller ranges.  For lower Rayleigh numbers we observed only two
stable patterns: dipole and hourglass.



\section{Conclusion}
\def\a#1{}
\a{Hof's configuration}

We have performed simulations with aspect ratio $\Gamma=2$ and Prandtl number
$Pr=6.7$, matching the configuration of Hof \etal{}~\cite{HofLucMul}, who
observed experimentally several different convective patterns at the same
Rayleigh number.  While their sidewalls were well insulating, we ran the
simulations for both perfectly conducting and perfectly insulating sidewalls.
The results of our simulations for insulating sidewalls are in good agreement
with the experiment: we succeeded in obtaining all five steady patterns
observed experimentally for $Ra=14\,200$.  For the same Rayleigh number we
observed the five stable steady solutions they reported: a toroidal roll; two,
three and four parallel rolls; and a three-spoke (``mercedes'') pattern.
Additionally, we obtained a new rotating S structure.

\a{Multiple states found, naming, summarising diagrams} 
For both types of boundary conditions we found multiple stable solutions for
the same Rayleigh number.  We preserved the nomenclature of Hof~\cite{HofThesis}
and gave names to some novel patterns.  We presented summary diagrams
organising the complicated dependencies between the coexisting stable
solutions.
\a{What we have done: confirmation of multiple states, even for low Ra} 
Our simulations confirm the fact that, even for cylinders of small aspect
ratio, the form of the convective flow depends not only on Rayleigh number,
but also dramatically on the initial condition. 
Furthermore, even for Rayleigh number as low as $2000$, 
we found up to three different long-lived patterns.

\a{Influence of boundary conditions}
The behaviour of the system depends both qualitatively and quantitatively on
the boundary conditions.  For higher Rayleigh numbers and perfectly insulating
boundary conditions, among the patterns we obtain are two, three and four
rolls, torus, mercedes and dipole smile.  When we change the sidewalls to
perfectly conducting, only three and four rolls can be observed. The
axisymmetric state disappears and the mercedes state is replaced by the Y
flow, thus losing its three-fold reflection symmetry. On the other hand, for
conductive sidewalls, new cross, da Vinci and star flows appear. The other patterns,
such as two rolls or dipole smile, may or may not also exist for conducting
sidewalls.  At lower Rayleigh numbers, insulating boundaries yield dipole,
pizza and axisymmetric patterns. For conducting boundaries the dipole flow
also exists, but the pizza pattern is replaced by hourglass and an
axisymmetric pattern appears only as a transient state.  In general, it seems
that switching to perfectly conducting sidewalls makes some of the patterns
lose their symmetries.  This transition could be elucidated by performing a
study in which intermediate values of sidewall conductivity would be used, and
checking how the flow symmetry changes.


\a{To be done: zoom in at the transition between high and low Ra}
Close to the threshold (at $Ra=2000$), for both types of boundary conditions,
we observed different stable patterns, but none of them seem to be stable at
higher Rayleigh number; conversely none of the patterns observed at higher
$Ra$ could be observed at $Ra=2000$. 
In a companion paper \cite{Boronska_PRE2}, we use Newton's method to 
construct a bifurcation diagram relating the states found at low $Ra$ to
those found at high $Ra$, for the case of insulating sidewalls.
In contrast to time-stepping, Newton's method converges more rapidly
and to states regardless of their stability.
The time-stepping procedure we used here lets us follow the dynamics known
from the experimental scenario, including transitions between the steady and
time-dependent states.  The patterns we obtained have constituted a 
good preliminary survey for constructing a complete bifurcation 
diagram via continuation. This task has not yet been undertaken 
for the case of conducting sidewalls.

\a{To be done: study of time-dependent states}

\a{What we did not find / Things to be done}
It is very likely that other stable solutions exist which we did not observe
because they were topologically too far from any of our initial conditions. 
We observed several new time-dependent flows, but we believe that we are far
from describing all existing unsteady solutions in the range of Rayleigh
numbers simulated. Except for the rotating S pattern which exists for
$Ra\approx 15\,000$, we observed oscillations mainly above $Ra=30\,000$.  
Hof \etal{}~\cite{HofLucMul} also observed a 3-loop rotating pattern at
$Ra=23\,000$ and a 13-spoke pulsing pattern at $Ra=33\,000$;
it would be interesting to reproduce these and to ascertain whether
they result from a Hopf bifurcation by a scenario like that 
described in our previous paper~\cite{Boronska_JFM}.
We showed the existence of several long-lasting transient states, for example
in the case of initialization with a dipole pattern and conducting boundaries,
for which the dipole smile state persisted for some time, with relatively
constant energy, before transforming into the final three-roll flow.  
This and other transitional patterns could also be objects for further study.
Despite the great variety of flows we have obtained, we are far from an
exhaustive study even for this specific configuration of control parameters.

\a{Closure}
Although small-aspect-ratio cylinders lack the universality of large ones, 
our three-dimensional numerical study has 
illustrated the remarkable variety of convective 
flows and rich dynamics found in this system.
We hope that our results convince the reader that this chapter of
hydrodynamical research is not yet thoroughly written.

\begin{acknowledgments}
We gratefully acknowledge the contributions of Daniel Benaquista to this work.  
We thank Tom Mullin and Bj\"orn Hof
for their continued interest in this work.  All of the computations were
performed on the computers of IDRIS (Institut pour le Developpement des
Ressources Informatiques et Scientifiques) of CNRS (Centre National pour la
Recherche Scientifique) under project 1119.
\end{acknowledgments}

\end{document}